# GPU-based Fast Cone Beam CT Reconstruction from Undersampled and Noisy Projection Data via Total Variation

**Xun Jia**

*Department of Radiation Oncology, University of California San Diego, La Jolla, CA 92037-0843*

**Yifei Lou**

*Department of Mathematics, University of California Los Angeles, Los Angeles, CA 90095*

**Ruijiang Li, William Y. Song, and Steve B. Jiang[a]**

*Department of Radiation Oncology, University of California San Diego, La Jolla, CA 92037-0843*

**Purpose**: Cone-beam CT (CBCT) plays an important role in image guided radiation therapy (IGRT). However, the large radiation dose from serial CBCT scans in most IGRT procedures raises a clinical concern, especially for pediatric patients who are essentially excluded from receiving IGRT for this reason. The goal of this work is to develop a fast GPU-based algorithm to reconstruct CBCT from undersampled and noisy projection data so as to lower the imaging dose.

**Methods:** The CBCT is reconstructed by minimizing an energy functional consisting of a data fidelity term and a total variation regularization term. We developed a GPU-friendly version of the forward-backward splitting algorithm to solve this model. A multi-grid technique is also employed.

**Results:** It is found that 20~40 x-ray projections are sufficient to reconstruct images with satisfactory quality for IGRT. The reconstruction time ranges from 77 to 130 sec on a NVIDIA Tesla C1060 GPU card, depending on the number of projections used, which is estimated about 100 times faster than similar iterative reconstruction approaches. Moreover, phantom studies indicate that our algorithm enables the CBCT to be reconstructed under a scanning protocol with as low as 0.1 mAs/projection. Comparing with currently widely used full-fan head and neck scanning protocol of ~360 projections with 0.4 mAs/projection, it is estimated that an overall 36~72 times dose reduction has been achieved in our fast CBCT reconstruction algorithm.

**Conclusions:** This work indicates that the developed GPU-based CBCT reconstruction algorithm is capable of lowering imaging dose considerably. The high computation efficiency in this algorithm makes the iterative CBCT reconstruction approach applicable in real clinical environments.

[a] Electronic mail: sbjiang@ucsd.edu

40     Cone Beam Computed Tomography (CBCT) has been broadly used in image guided radiation therapy (IGRT) to acquire the updated patient's geometry for precise targeting before each treatment fraction. The repeated use of CBCT during a treatment course raises a clinical concern of excessive x-ray dose. This concern has prohibited the use of IGRT for pediatric patients, resulting in compromised treatment outcome.

45     Imaging dose in CBCT can be reduced by reducing number of x-ray projections and/or mAs level (tube current and pulse duration). These approaches cannot be used with conventional FDK-type algorithms[1], that are currently clinical standards, because the images reconstructed from under-sampled and/or noisy projection data are highly degraded and thus clinically unacceptable. Recently, a burst of research in compressed
50 sensing[2-3] has demonstrated the feasibility of recovering signals from incomplete measurements through optimization methods, providing us new perspectives of solving the CBCT reconstruction problem[4]. Among various methods of this type, Total Variation (TV) method[5] has presented its tremendous power in CT reconstruction problems in both fan-beam[6] and cone-beam[7] geometries. This approach has also been extensively applied
55 into many other imaging applications[8-10] and its efficacy has been enhanced by combining with other techniques, such as incorporating prior information[11]. Despite the great power of the TV-based methods, the computation is very time-consuming owing to the lack of efficient algorithms to handle the large data set encountered in CBCT reconstruction problems. It usually takes several hours or even longer for current TV-based
60 reconstruction approaches to produce a decent CBCT image. This fact prevents them from practical applications in real clinical environments.

    Recently, general-purpose computing on graphics processing unit (GPU) has offered us a promising prospect of performing computationally intensive tasks in medical imaging and therapy applications[12-17]. By developing new algorithms with mathematical
65 structures suitable for GPU parallelization, we can take advantage of the massive computing power of GPU to dramatically improve the efficiency of the TV-based CBCT reconstruction, as will be seen in the rest of this letter.

    Let us consider a patient volumetric image represented by a function $f(x,y,z)$, $(x,y,z) \in \mathbf{R}^3$. An operator $P^\theta$ projects $f$ onto an x-ray imager plane in a cone beam
70 geometry at an angle $\theta$. Denote the observed projection image at an angle $\theta$ by $Y^\theta(u,v)$. A CBCT reconstruction problem is formulated as to retrieve the volumetric image $f$ based on the observed functions $Y^\theta$. In this letter, we aim at reconstructing the CBCT image by minimizing an energy functional:

$$f = \mathrm{argmin}\, E[f] = \mathrm{argmin}\, E_1[f] + \mu E_2[f],$$
$$s.t.\ f(x,y,z) \geq 0 \text{ for } \forall\, (x,y,z) \in \mathbf{R}^3, \quad (1)$$

where $E_1[f] = \frac{1}{V}\|\nabla f\|_1$ and $E_2[f] = \frac{1}{N_\theta A}\sum_\theta \|P^\theta[f] - Y^\theta\|_2^2$. Here $V$ is the volume in
75 which the CBCT image is reconstructed. $N_\theta$ is the number of projections and $A$ is the projection area on each x-ray imager. $\|...\|_p$ denotes function $l_p$-norm. The data fidelity term $E_2$ ensures the consistency between the reconstructed image $f$ and the observations $Y^\theta$. The other one, $E_1$, known as TV semi-norm, has been shown to be extremely powerful[5] to remove artifacts and noise from $f$ while preserving its sharp edges. $\mu > 0$ is 80   introduced to adjust the relative weights between these two energy terms. It controls the smoothness of the reconstructed images and is chosen manually for best image quality.

In order to perform the minimization task, we employed an innovative forward-backward splitting algorithm[18-19]:

**Algorithm A1:**

Repeat the following steps until convergence

1. Update: $g = f - \frac{\mu V}{\beta} \frac{\delta}{\delta f} E_2[f]$;
2. Minimize: $f = \mathrm{argmin}\, E_1[f] + \frac{\beta}{2} E_3[f]$;
3. Correct: $f(x, y, z) = 0$, if $f(x, y, z) < 0$.

where $\beta > 0$ is a parameter and $g(x, y, z)$ is an auxiliary function. The choice of $\beta$ only
85  affects the stability of the algorithm and an empirical value $\beta \sim 10\mu$ is used in our implementation. The energy functional in Step 2 is defined as $E_3[f] = \frac{1}{V}\|f - g\|_2^2$ and the sub-problem in this step is solved by a simple gradient descent method.

A straightforward way of implementing (A1) is to interpret $P^\theta[f]$ as a matrix multiplication and then $E_2[f]$ as a matrix norm $\sum_\theta \|P^\theta f - Y^\theta\|_2^2$. This leads to a simple
90  form $\sum_\theta {P^\theta}^T (P^\theta f - Y^\theta)$ for the functional variation $\delta E_2[f]/\delta f(x,y,z)$ in Step 1, apart from some constants, where $\cdot^T$ denotes a matrix transpose. This approach causes a memory conflict problem when implemented on GPU, which severely limits the exploitation of GPU's massive parallel computing power. To resolve this issue, a GPU-friendly way of evaluating the functional variation $\delta E_2[f]/\delta f$ has been developed as
95  following:

$$\frac{\delta}{\delta f(x,y,z)} E_2[f] = \frac{1}{N_\theta A} \sum_\theta \frac{2L^3(u^*,v^*)}{L_0 l^2(x,y,z)} \cdot [P^\theta[f](u^*,v^*) - Y^\theta(u^*,v^*)], \quad (2)$$

where $(u^*, v^*)$ is the coordinate on the imager plane at which a ray line connecting the x-ray source and the point $(x, y, z)$ intersects the imager. $L_0$ is the source to imager distance. $l(x, y, z)$ and $L(u^*, v^*)$ are the distances from the source to the point $(x, y, z)$ and from the source to the point $(u^*, v^*)$ on the imager, respectively. For GPU
100  implementation, we first perform the forward x-ray projection operation and compute $[P^\theta[f](u,v) - Y^\theta(u,v)]$ for all $(u,v)$ and $\theta$. Then each thread on GPU can independently evaluate the functional variation at a $(x, y, z)$ coordinate. Given the vast parallelization ability of GPU, extremely high efficiency in Step 1 can be achieved. Other steps in (A1) can also benefit from GPU implementations considerably. For example, for
105  the purpose of evaluating $E_1[f]$, we simply have each GPU thread compute this term at an $(x, y, z)$ coordinate and then a summation over the coordinate is performed. Moreover, we employed the well developed multi-grid method[20] to achieve further efficiency boost.

We first tested our reconstruction algorithm on a digital NCAT phantom[21]. The phantom was generated at thorax region with a size of $512 \times 512 \times 70$ voxels. The x-ray
110  imager was modeled to be an array of $512 \times 384$ detectors. X-ray projections of the phantom were generated along various directions and were then used as the input for the CBCT reconstruction. Axial slices of the reconstruction results based on $N_\theta = 20, 30$ and

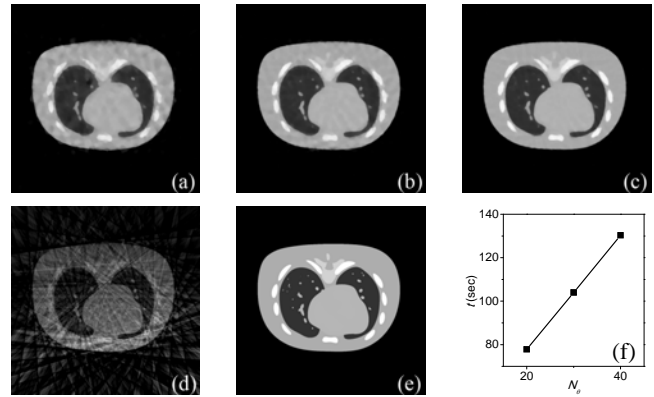

**FIG. 1**. (a) to (c) Reconstructed CBCT images under $N_\theta =$ 20, 30, and 40 x-ray projections. (d) Reconstruction result from the FDK algorithm under 40 projections. (e) The ground-truth image. (f) Reconstruction time $t$ as a function of $N_\theta$.

40 projections are drawn in Fig. 1(a) through 1(c), respectively. In all cases, the projections are taken along equally spaced angles covering an entire 360 degree rotation. For the purpose of comparison, we also show in panels (d) and (e) the image reconstructed from conventional FDK algorithm[1] as well as the ground-truth image. Clearly, the reconstructed CBCT images based on 40 projections is almost visually indistinguishable from the ground-truth. On the other hand, the image produced by the conventional FDK algorithm is full of streaking artifacts due to the insufficient number of projections. Moreover, the required number of projections can be further lowered depending on the clinical purposes. For example, 20 projections may suffice for patient setup purposes in radiotherapy, where major anatomical features have already been retrieved as in Fig. 1(a). As far as radiation dose is concerned, the results shown in Fig. 1 implies a 9~18 times dose reduction comparing with the currently widely used FDK algorithm, where about 360 projections are usually taken in a full-fan head and neck scanning protocol. More importantly, we emphasize here that the total reconstruction time is short enough for real clinical applications. In Fig. 1(f), we plot the dependence of the reconstruction time $t$ as a function of the number of projections $N_\theta$. As we can see, the reconstructions can be accomplished within 77~130 seconds on an NVIDIA Tesla C1060 GPU card, depending on the number of projections. Comparing with the computation time of several hours in currently similar reconstruction approaches, our algorithm has achieved a tremendous efficiency enhancement (~100 times speed up).

To demonstrate our algorithm's capability of handling noisy data, we performed CBCT reconstruction on a CatPhan 600 phantom (The Phantom Laboratory, Inc., Salem, NY) under different mAs levels. 379 projections within 200 degrees were acquired by Varian On-Board Imager system (Varian Medical Systems, Palo Alto, CA). A subset of equally spaced 40 projections was used to perform the reconstruction. In Fig. 2, we show the reconstruction results based on various mAs levels using our TV-based algorithm and the FDK algorithm. Again, the images produced by our method are smooth and free of streaking artifacts, undoubtedly outperforming those from the FDK algorithm. In particular, under an extremely low mAs level (0.1 mAs/projection), our method is still able to capture major features of the phantom. Comparing with the currently widely used full-fan head and neck scan protocol of 0.4 mAs/projection, this performance implies a dose reduction by a factor of ~4 due to lowering the mAs level. Taking into account the dose reduction by reducing the number of x-ray projections, an overall 36~72 times dose reduction has been achieved.

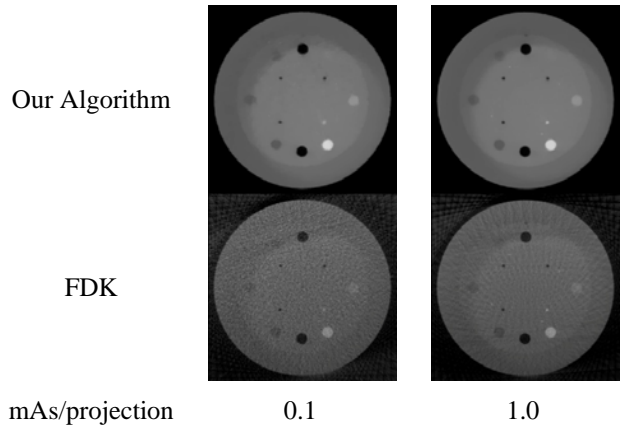

| Our Algorithm | | |
| FDK | | |
| mAs/projection | 0.1 | 1.0 |

**FIG. 2**. Axial slices in the CBCT reconstruction results of a CatPhan 600 phantom under two different mAs levels from FDK algorithm and from our algorithm. 40 projections are used in all cases.

In this letter, we presented our recent development on a fast iterative algorithm for the CBCT reconstruction problem. We consider an energy functional consisting of a data fidelity term and a regularization term of TV semi-norm. The minimization problem is solved with a forward-backward splitting method together with a multi-grid approach on a GPU platform, leading to both satisfactory accuracy and efficiency. Reconstruction performed on a digital NCAT phantom indicates that images with decent quality can be reconstructed from 20~40 x-ray projections and the total reconstruction time ranges from 77 to 130 seconds depending on the number of projections used. We have also tested our algorithm on a CatPhan 600 phantom under different mAs level and found that CBCT images can be successfully reconstructed from scans with as low as 0.1 mAs/projection. All of these results indicate that our new algorithm has improved the efficiency by a factor of 100 over existing iterative algorithms and reduced imaging dose by a factor of 36~72 compared to the current clinical standard full-fan head and neck scanning protocol. The high computation efficiency achieved in our algorithm makes the iterative CBCT reconstruction approach applicable in real clinical environments for the first time.

This work is supported in part by the University of California Lab Fees Research Program. We would like to thank NVIDIA for providing GPU cards for this project.

## References

1. L. A. Feldkamp, L. C. Davis and J. W. Kress, "Practical Cone-Beam Algorithm," J. Opt. Soc. Am. A-Opt. Image Sci. Vis. **1**, 612-619 (1984).
2. E. J. Candes and T. Tao, "Near-optimal signal recovery from random projections: Universal encoding strategies?," Ieee Transactions on Information Theory **52**, 5406-5425 (2006).
3. D. L. Donoho, "Compressed sensing," Ieee Transactions on Information Theory **52**, 1289-1306 (2006).
4. J. Wang, T. F. Li and L. Xing, "Iterative image reconstruction for CBCT using edge-preserving prior," Medical Physics **36**, 252-260 (2009).
5. L. I. Rudin, S. Osher and E. Fatemi, "Nonlinear total variation based noise removal algorithms," Physica D **60**, 259-268 (1992).
6. E. Y. Sidky, C. M. Kao and X. H. Pan, "Accurate image reconstruction from few-views and limited-angle data in divergent-beam CT," J. X-Ray Sci. Technol. **14**, 119-139 (2006).
7. E. Y. Sidky and X. C. Pan, "Image reconstruction in circular cone-beam computed tomography by constrained, total-variation minimization," Phys. Med. Biol. **53**, 4777-4807 (2008).
8. M. Persson, D. Bone and H. Elmqvist, "Three-dimensional total variation norm for SPECT reconstruction," Nuclear Instruments & Methods in Physics Research Section a-Accelerators Spectrometers Detectors and Associated Equipment **471**, 98-102 (2001).
9. E. Y. Sidky, X. C. Pan, I. S. Reiser, R. M. Nishikawa, R. H. Moore and D. B. Kopans, "Enhanced imaging of microcalcifications in digital breast tomosynthesis through improved image-reconstruction algorithms," Medical Physics **36**, 4920-4932 (2009).
10. H. Y. Yu and G. Wang, "Compressed sensing based interior tomography," Physics in Medicine and Biology **54**, 2791-2805 (2009).
11. G. H. Chen, J. Tang and S. H. Leng, "Prior image constrained compressed sensing (PICCS): A method to accurately reconstruct dynamic CT images from highly undersampled projection data sets," Medical Physics **35**, 660-663 (2008).
12. F. Xu and K. Mueller, "Accelerating popular tomographic reconstruction algorithms on commodity PC graphics hardware," Ieee Transactions on Nuclear Science **52**, 654-663 (2005).
13. F. Xu and K. Mueller, "Real-time 3D computed tomographic reconstruction using commodity graphics hardware," Physics in Medicine and Biology **52**, 3405-3419 (2007).
14. X. Gu, D. Choi, C. Men, H. Pan, A. Majumdar and S. Jiang, "GPU-based ultra fast dose calculation using a finite size pencil beam model " Phys. Med. Biol. **54**, 6287-6297 (2009).
15. C. Men, X. Gu, D. Choi, A. Majumdar, Z. Zheng, K. Mueller and S. Jiang, "GPU-based ultra fast IMRT plan optimization," Phys. Med. Biol. **54**, 6565-6573 (2009).
16. X. J. Gu, H. Pan, Y. Liang, R. Castillo, D. S. Yang, D. J. Choi, E. Castillo, A. Majumdar, T. Guerrero and S. B. Jiang, "Implementation and evaluation of various demons deformable image registration algorithms on a GPU," Physics in Medicine and Biology **55**, 207-219 (2010).
17. X. Jia, X. Gu, J. Sempau, D. Choi, A. Majumdar and S. B. Jiang, "Development of a GPU-based Monte Carlo dose calculation code for coupled electron-photon trasport," Physics in Medicine and Biology, arXiv:0910.0329 (2010).
18. P. L. Combettes and V. R. Wajs, "Signal recovery by proximal forward-backward splitting," Multiscale Modeling & Simulation **4**, 1168-1200 (2005).
19. E. T. Hale, W. T. Yin and Y. Zhang, "Fixed-point continuation for l1-minimization: methodology and convergence," Siam Journal on Optimization **19**, 1107-1130 (2008).
20. A. Brandt, "Multiscale scientific computation: review," in *Multiscale and multiresolution methods: theory and applications,* edited by T. J. Barth, T. F. Chan and R. Haimes (Springer, 2002), pp. 53.
21. W. P. Segars, D. S. Lalush and B. M. W. Tsui, "A realistic spline-based dynamic heart phantom," Ieee Transactions on Nuclear Science **46**, 503-506 (1999).